\documentclass[PRL,twocolumn,showpacs,preprintnumbers,amsmath,amssymb,superscriptaddress]{revtex4}

\usepackage{graphicx}
\usepackage{epsfig}
\usepackage{amsmath}
\usepackage{amssymb}
\usepackage{textcomp}
\usepackage{dcolumn}
\usepackage{bm}

\newcommand{\mm}{\textrm{mm}}

\newcommand{\mK}{\textrm{mK}}
\newcommand{\mV}{\textrm{mV}}
\newcommand{\eV}{\textrm{eV}}
\newcommand{\kV}{\textrm{kV}}
\newcommand{\nanom}{\textrm{nm}}

\newcommand{\nanos}{\textrm{ns}}
\newcommand{\V}{\textrm{V}}

\begin{document}
\newcommand*{\ULM}{Institut f\"ur Quanteninformationsverarbeitung,
Universit\"at Ulm, Albert-Einstein-Allee 11, 89081 Ulm, Germany}
\affiliation{\ULM}
\homepage{http://www.quantenbit.de}

\newcommand*{\BOCHUM}{RUBION, Ruhr-Universit\"at Bochum,
44780 Bochum, Germany}
\affiliation{\BOCHUM}

\title{Deterministic Ultracold Ion Source targeting the Heisenberg Limit}
\author{W. Schnitzler}\affiliation{\ULM}
\author{N. M. Linke}\affiliation{\ULM}
\author{R. Fickler}\affiliation{\ULM}
\author{J. Meijer}\affiliation{\BOCHUM}
\author{F. Schmidt-Kaler}\affiliation{\ULM}
\author{K. Singer}\email{kilian.singer@uni-ulm.de}\affiliation{\ULM}


\date{\today}

\begin{abstract}

The major challenges to fabricate  quantum processors and future
nano solid state devices are material modification techniques with
nanometre resolution and suppression of statistical fluctuations
of dopants or qubit carriers. Based on a segmented ion trap with
\mK~laser cooled ions we have realized a deterministic single ion
source which could operate with a huge range of sympathetically
cooled ion species, isotopes or ionic molecules. We have
deterministically extracted a predetermined number of ions on
demand and have measured a longitudinal velocity uncertainty of
6.3m/s and a spatial beam divergence of 600$\mu$rad. We show in
numerical simulations that if the ions are cooled to the motional
ground state (Heisenberg limit) nanometre spatial resolution can
be achieved.

\end{abstract}


%
%
%
%

\maketitle

The miniaturization of semiconductor devices has reached length
scales of a few tens of nanometres, where statistical Poissonian
fluctuations of the number of doping atoms in a single transistor
significantly affect the characteristic properties, e.g. gate
voltage or current amplification \cite{SHINADA2005}. Further
miniaturization will even cause statistical device failure.
Particularly fatal are statistical dopant fluctuations for a
future solid state quantum processor based on single implanted
qubit carriers like colour centres in diamond or phosphorous
dopants in silicon
\cite{GURUDEV2007,NEUMANN2008,KANE1998,GREENTREE2008}. So far, the
only known methods to control the number of dopants utilize
statistical thermal sources followed by a post-detection of the
implantation event, either by the observation of Auger electrons,
photoluminescence, phonons, the generation of electron-hole pairs
or changes in the conductance of field effect transistors
\cite{SHINADA2002,PERSAUD2004,MITIC2005,BATRA2007,SHINADA2008}. To
make the detection of such an event successful the methods require
either highly charged ions or high implantation energies which, as
a down side, generate defects in the host material. In these
systems resolutions of less than 10\nanom~are achieved by means of
masks and apertures shielding the substrate from incident ions and
leading to compulsory losses of dopants. Another fabrication
method, specific for Si-surfaces, uses hydrogen terminated
surfaces structured with the tip of a tunneling microscope,
followed by a chemical reactive surface binding of doping atoms
\cite{OBRIEN2001,SCHOFIELD2003,RUESS2004,POK2007,RUESS2007}. With
this technique sub \nanom~resolution can be achieved but the
applicability is mainly limited to specific substrates and
impurities in the background gas can cause severe impairment.

\begin{figure}[h]
\includegraphics[width=\columnwidth]{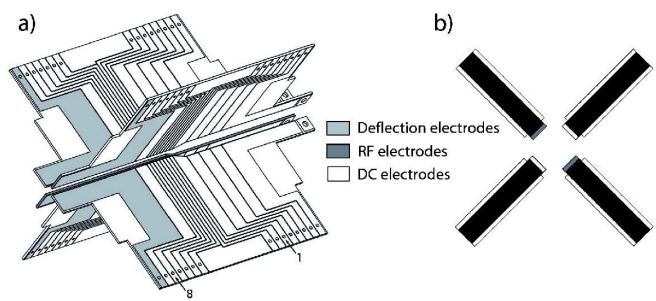}
\caption{a) Sketch of the segmented linear Paul trap with
DC-electrodes (white) and RF-electrodes (dark grey).
Deflection electrodes (light grey) are used to alter
the trajectories of ions which are extracted out of the trap. b)
Front view showing that RF and RF-ground electrodes (generating the radial confinement) 
are only covering the two 410$\mu$m~wide front faces of the blades.}
\label{fig:fig1}
\end{figure}

Here, we present the experimental proof of a novel ultracold ion
source which can be used for the deterministic implantation of a
predetermined number of single ions~\cite{MEIJER2006}. Our
technique is based on a segmented linear Paul trap with laser
cooled $^{40}\mathrm{Ca}^+$ ions similar to setups used for
scalable quantum information processing with ions \cite{ROWE2002}.
Additionally loaded doping ions of different elements or ionic
molecules cannot be directly laser cooled but could be
sympathetically cooled by $^{40}\mathrm{Ca}^+$ ions. Although
invisible to the laser light they are still identified
\cite{DREWSEN2004,NAEGERL1998} and counted by exciting collective
vibrational modes. Our segmented ion trap allows for the
separation of the cooling ion from the dopant ion, which is
finally extracted by a tailored electric field. The implantation
method is in principle independent of the dopant species and the
target substrate. For 2\mK~laser-cooled ions accelerated to
80\eV~the measured longitudinal velocity distribution shows a
1$\sigma$-spread of 6.3(6)m/s \cite{LEE1998} and a spatial
1$\sigma$-spot radius of 83($^{+8}_{-3}$)$\mu$m~at a distance of
257\mm~(beam divergence: 600$\mu$rad). These properties reduce
chromatic and spherical aberration of any focusing ion optics. The
resolution of our system is thereby not enforced by additional
masks or apertures but is an intrinsic property of our setup.

The core of the experimental setup is a Paul trap - a universal
tool for trapping charged particles such as atomic and molecular
ions or charged clusters using a combination of static (DC) and
alternating (RF) electric fields. A pseudo-potential of a few eV
depth is generated with a properly chosen RF amplitude and
frequency $\Omega$.

\begin{figure}[htb]
\includegraphics[width=0.85\columnwidth]{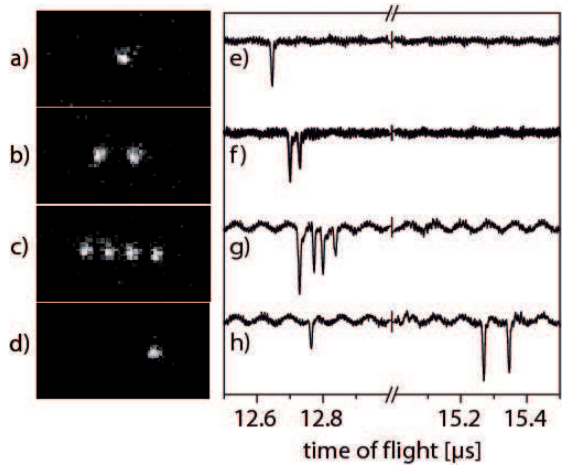}
\caption{Typical fluorescence image of a single
$^{40}\mathrm{Ca}^+$ ion (a), and linear ion crystals of two (b)
and four (c) ions. After the extraction we record EMT detector
signal traces with a single ion detection event (e), two events
(f) and four events (g), correspondingly. The EMCCD image (d)
shows fluorescence from a single $^{40}\mathrm{Ca}^+$ ion only;
however, we can discover from its position that it is trapped in a
linear crystal together with two dark ions at the left hand side.
As the mixed ion crystal is extracted, we detect three events, one
from the $^{40}\mathrm{Ca}^+$ near 12.8$\mu$s and two events near
15.3$\mu$s. From this time-of-flight spectroscopy, we reveal the
mass of $\mathrm{CaO}^+$ ions for both dark ions. All measurements
were conducted without the movable aperture plate with an
effective distance of 247\mm~between trap centre and detector.}
\label{fig:fig2}
\end{figure}

For our application it is necessary that the ions arrange as a
linear crystal such that they can be identified and counted using
laser induced fluorescence. During the extraction we apply
voltages to additional DC segments tailoring the axial potential.
In a conventional linear segmented Paul trap this would lead to a
loss of the radial confinement because the applied extraction
potential exceeds the radial pseudo-potential. We have developed a
special design of our trap, in which the ions are radially guided
even during the axial extraction. Our trap consists of four copper
plated polyimide blades of 410$\mu$m thickness and 65\mm~length
which are arranged in a x-shaped manner \cite{HUBER2008} (see
fig.~\ref{fig:fig1} for a schematic view). The RF is applied to
the inner front faces of two opposing blades; the front faces of
both other blades are grounded. The distance between inner front
faces of opposing blades is 2\mm. DC voltages are applied to eight
segments of 0.7\mm~width which are placed on the top and bottom
areas of all four blades. Under typical operating conditions we
apply to the RF electrodes an amplitude of 200\V~at the frequency
of $\Omega/2\pi$ = 12.155MHz leading to a radial secular frequency
$\omega_\mathrm{rad.}/2\pi$ = 430kHz for a $^{40}\mathrm{Ca}^+$
ion. The DC-electrode trap segments 2 and 8 are supplied with
35\V~and the remaining electrodes with 0\V~resulting in an axial
potential with $\omega_\mathrm{ax.}/2\pi$ = 280kHz. The location
of trapped ions is above electrode 5. The trap assembly is housed
in a stainless steel vacuum chamber with enhanced optical access
held by a turbomolecular pump and an ion-getter pump at a pressure
of 3$\times10^{-9}\text{mbar}$. Ions are illuminated by resonant
laser light near 397nm and 866nm for Doppler cooling. Scattered
photons are collected by a f/1.76 lens on a EMCCD camera to image
individual $^{40}\mathrm{Ca}^+$ ions, see fig.~\ref{fig:fig2}(a) -
(c). From the width of the laser excitation spectrum on the
$S_\mathrm{1/2}$ - $P_\mathrm{1/2}$ laser cooling transition, we
deduce a temperature of about 2\mK~slightly above the Doppler
cooling limit.

\begin{figure}[htb]
\includegraphics[width=0.80\columnwidth]{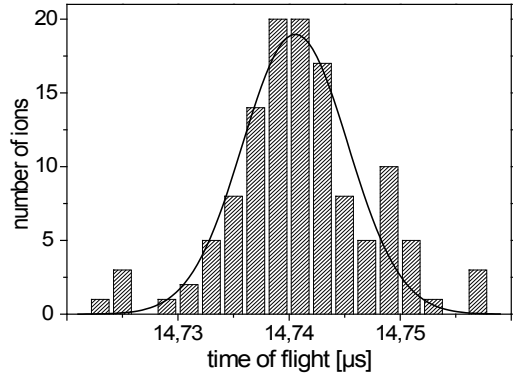}
\caption{Time-of-flight distribution for single ions based on 123 successful extractions out of 139 shots in total through the 1\mm~aperture. The bin size of the histogram is 2\nanos. A Gaussian fit of the data yields an average velocity of 19.47km/s with a 1$\sigma$-spread of 6.3(6)m/s.}
\label{fig:fig3}
\end{figure}

Calcium and dopant ions are generated in a multi-photon ionization
process by a pulsed frequency tripled Nd-YAG laser at 355nm with a
pulse power of 7mJ. Dopant ions are sympathetically cooled and
identified from the voids in the fluorescence image compared to
that of a regular linear $^{40}\mathrm{Ca}^+$ crystal.
Fig.~\ref{fig:fig2}(d) shows the fluorescence of an ion crystal
consisting of a single $^{40}\mathrm{Ca}^+$ and two molecular
$\mathrm{CaO}^+$ ions resulting from a chemical reaction with
background residual gas \cite{DREWSEN2004}. We identify the
species of dark ions by exciting collective vibrational modes with
an AC voltage applied to electrode 4 and observing a blurring of
the $^{40}\mathrm{Ca}^+$ fluorescence image at the resonance
frequency $\omega_\mathrm{ax.}$ \cite{NAEGERL1998}. Alternatively,
amplitude modulated resonant laser light is used
\cite{DREWSEN2004}~to determine the charge to mass ratio of
trapped particles at a precision of better than 0.2\%.  Before
extraction, the sympathetically cooled doping ions may be
separated from the $^{40}\mathrm{Ca}^+$ ions. This is achieved by
converting the axial trapping potential into a double well. The
doping ions are further transported away from the
$^{40}\mathrm{Ca}^+$ ions by time dependent DC electrode voltages
\cite{HUBER2008}. As heating generated during this separation
process \cite{ROWE2002} cannot be cooled away anymore an
alternative separation method would deflect the unwanted
$^{40}\mathrm{Ca}^+$ ions after extraction e.g. by increasing the
electrode voltages of an einzel-lens. For the extraction we
increase the DC voltage of segments 4 and 5 to 500\V~within a few
tens of nanoseconds. The switching of the extraction voltage
(supplied by iseg inc., Model EHQ-8010p) is performed by two high
voltage switches (Behlke inc., HTS 41-06-GSM) triggered via a
computer-controlled TTL-signal and synchronized with the RF-field
phase. Synchronization is crucial in order to minimize shot to
shot fluctuations of velocity and position. An electronic phase
synchronization circuit delays the TTL signal for extraction such
that a constant delay to the next zero-crossing of the trap drive
with frequency $\Omega$ is ensured. We found the optimum
extraction parameters by matching the time of extraction with a
certain phase of the radio frequency and by adjusting the
DC-voltages on the deflection electrodes, which alter the ion
trajectory during extraction. All measurements described below use
these settings. The detection of the extracted ions is performed
via an electron multiplier tube (EMT) with 20 dynodes from ETP
inc., Model AF553, which can detect positively charged ions with a
specified quantum efficiency of about 80\%. The detector is housed
in a separate vacuum chamber at a distance of 287\mm~from the
trap. At typical operating conditions the detector is supplied
with a voltage of -2.5\kV. The gain is specified with
5$\times10^{5}$ and we observe an electrical signal of about
100\mV. The detection events show a width of 10 to 15\nanos. In
order to measure the beam divergence a movable aperture plate was
installed in front of the detector. This plate, mounted on a
nanopositioning stage from Smaract, Model SL-2040, features hole
diameters ranging from 5\mm~down to 300$\mu$m.

Typical EMT detector signals for different numbers of ions are
shown in fig.~\ref{fig:fig2}(e) - (g). Fig.~\ref{fig:fig2}(h)
displays the detector events for one $^{40}\mathrm{Ca}^+$ ion and
two $\mathrm{CaO}^+$ ions, which arrive at t=15.3$\mu$s. From a
time-of-flight analysis through the 1\mm~aperture we deduce a mean
ion velocity of 19.47km/s for the $^{40}\mathrm{Ca}^+$ ions. At
3$\times10^{-9}\text{mbar}$ we detect 87($^{+7}_{-11}$)\% of all
extracted single ions within a 1$\sigma$-confidence interval. We
found that the efficiency slightly depends on the residual gas
pressure but is mainly limited by the detector efficiency (which
we measure to be higher than specified). The measured longitudinal
velocity distribution (see fig.~\ref{fig:fig3}) shows a
1$\sigma$-spread of 6.3(6)m/s which is about a factor of 10 larger
than the velocity distribution inside the trap at T=2\mK. This
leads to a relative velocity uncertainty $\Delta$v/v of 3.2$\times
10^{-4}$ which may be further reduced by post-accelerating the
ions after extraction. From measurements conducted with the
smallest aperture (300$\mu$m) we deduce a 1$\sigma$-spot radius of
83($^{+8}_{-3}$)$\mu$m~for the trajectories of the extracted ions.
Here we assume a Gaussian spatial distribution and the error is
due to counting statistics. Note that this value is an upper limit
as our measurements are currently affected by a measured drift of
the ion beam of about 15$\mu$m/min possibly caused by temperature
drifts of the setup.

\begin{figure}[htb]
\includegraphics[width= 0.9\columnwidth]{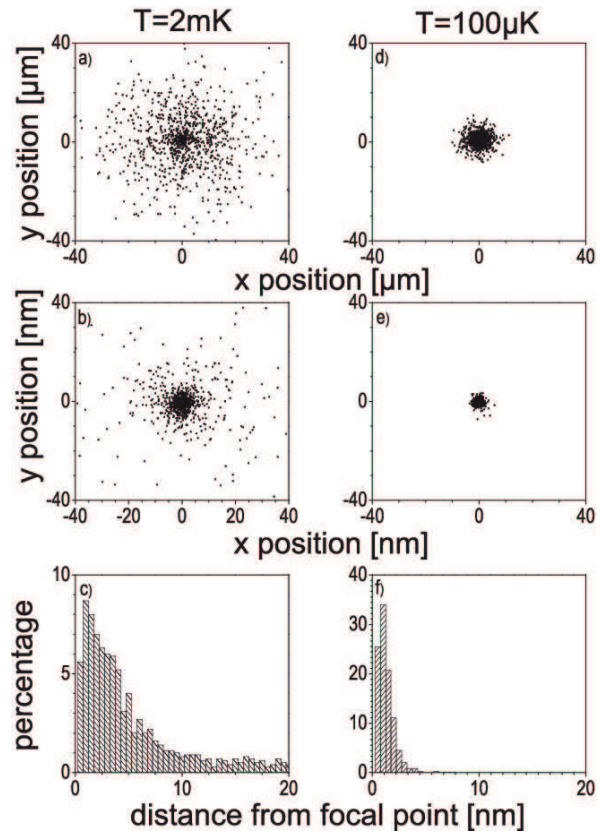}
\caption{Monte-Carlo simulation of extracted ions. Left side: (a) Spot diagram at a distance of 247\mm~from the trap centre for an initial ion temperature of 2\mK. (b) Focal spot diagram generated by an einzel-lens with a 1$\sigma$-spot radius of 7\nanom. (c) Histogram of radial distribution of spots in the focal plane. Right side: (d)-(f) Similar diagrams as on the left side but for a temperature of 100$\mu$K.}
\label{fig:fig4}
\end{figure}

For a comparison of measured data with numerical Monte-Carlo
simulations we need accurate electrostatic potentials which we
deduce from a complete CAD-model of the trap geometry created with
AutoCAD. Electrostatic potentials and fields are calculated by
using a boundary element method accelerated by the fast multipole
method \cite{GREENGARD1988}. Symmetry properties of the trap are
exploited to reduce numerical errors. The ion trajectories are
obtained by applying the Verlet integration method. The initial
momentum and position is determined from the thermal Boltzmann
distribution in the trapping potential. As a test, we have
compared measured trap frequencies $\omega_\mathrm{ax.}$ and
$\omega_\mathrm{rad.}$ for various traps of different size and
shape in our lab \cite{SCHULZ2008}~with corresponding simulations
and found an agreement at the level of 2 to 3\%. The ion
trajectory calculation takes into account the full time dependent
dynamics, including the micro-motion at frequency $\Omega$ yielding a
1$\sigma$-velocity spread of 12m/s and a beam divergence of
130$\mu$rad. Simulated velocity uncertainty and beam divergence agree within
one order of magnitude with experimental results (see tab.
\ref{tab:tab1}).

In order to implant single ions into solid state materials with
nanometre spatial resolution, the detector will be replaced by a simple
electrostatic einzel-lens \cite{SEPTIER1960}~with a diameter of
1\mm~and a focal length of 9\mm. Simulations predict a
1$\sigma$-spot radius of 7\nanom~for 2\mK~and 2\nanom~for
100$\mu$K respectively (see fig.~\ref{fig:fig4} and
tab.~\ref{tab:tab1}).

\begin{table}[htb]
\begin{tabular}{l|cccc}
& T & $\Delta$v & $\alpha$ & $r_f$\\
\hline
meas. & 2mK & 6.3m/s & 600$\mu$rad & -\\
calc. & 2mK & 12m/s & 130$\mu$rad & 7nm \\
calc. & 100$\mu$K & 1m/s & 30$\mu$rad & 2nm \\
\end{tabular}
\caption{Comparison between experimental and numerical 1-$\sigma$
longitudinal velocity uncertainty $\Delta$v, beam divergence
$\alpha$ (full angle) and 1-$\sigma$ focal spot radius $r_f$ for
different initial ion temperatures $T$.} \label{tab:tab1}
\end{table}

We attribute the discrepancies between experimental results and
numerical simulations to patch electric fields on insulating
surfaces, geometrical imperfections of the electrodes and
fluctuations of the extraction voltage power supply (specified
with  $\Delta$U/U = $10^{-5}$). The aforementioned drift of the
ion beam will be reduced in future experiments by installing
magnetic shielding and a temperature stabilization.

Currently, our inital mean spatial and momentum spread is a factor
of 10 larger than at the Heisenberg limit. An ideally suited
cooling method for reaching this fundamental limit uses the
electromagnetically induced transparency, as it allows cooling of
all degrees of freedom at different oscillation frequencies even
for mixed ion crystals \cite{ROOS2000,KALER2001}. This would lead
to the perfect single-ion single-mode matter-wave source. By
changing the trapping parameters we can freely adjust the ratio
between the variance of the spatial components versus variance of
the momentum components.

Thus, the spot size would be limited by the diffraction of the
matter wave, which results in a spot size of $10^{-10}$m if we
assume a numerical aperture of 0.001 for the ion lens and an
energy of 80\eV. To assure the proper alignment of a short focal
length lens-system with respect to the substrate we propose to
implant through a hole in the tip of an atomic force microscope
\cite{PERSAUD2005,MEIJER2008}. This would bring along the
additional advantage that the charged particle could be placed
with respect to surface structures (such as gate electrodes) in
deterministic doping applications. A further possible application
of our system is as on-demand source for matter wave
interferometry with ultracold slow ions, which until now was only
possible with electrons, neutrons and neutral atoms and molecules
\cite{JOENSSON1961,CARNAL1991,ARNDT1999,RAUCH1974}. Being
compatible with state of the art ion trap quantum processors our
setup may be used to convey qubits directly from one trap to the
other by transmitting the qubit carrier itself.

In conclusion we have experimentally realized a deterministic
ultracold source for single ions and ionic molecules. For an ion
temperature of a few mK we measured a longitudinal velocity
distribution of extracted ions which shows a 1$\sigma$-spread of a
few meters per second which is a promising starting point for the
application of ion optical elements. Ion ray tracing simulations
predict nm resolution for our setup when combined with an
electrostatic einzel-lens. If the ions are further cooled to the
motional ground state our setup could realize the perfect matter
wave source at the Heisenberg limit.

We acknowledge financial support by the Landes\-stiftung
Baden-W\"urttemberg in the framework 'atomics' (Contract No. PN
63.14), the European commission within EMALI (Contract No.
MRTN-CT-2006-035369) and the Volkswagen Stiftung.

\vspace{1cm}

\end{document}